\documentclass[letterpaper]{article}
\usepackage{aaai2026}
\usepackage{times}
\usepackage{helvet}
\usepackage{courier}
\usepackage[hyphens]{url}
\usepackage{graphicx}
\usepackage{natbib}
\usepackage{caption}
\usepackage{xcolor}
\usepackage{tikz}
\usetikzlibrary{calc,arrows.meta,backgrounds,positioning}
\usepackage{polycouncil}

\nocopyright

\newcommand{\sysname}{KITA AI}

\title{KITA AI: A Multi-Agent LLM System for Pluralistic Policy Deliberation}
\author{
    Arnau Mayoral-Macau\textsuperscript{\rm 1}\equalcontrib,
    Jiaqi Lai\textsuperscript{\rm 2}\equalcontrib,
    Manala Tyobeka\textsuperscript{\rm 3},
    Vukosi Marivate\textsuperscript{\rm 3},\\
    William Chandra Tjhi\textsuperscript{\rm 2},
    Georgina Curto\textsuperscript{\rm 1}
}
\affiliations{
    \textsuperscript{\rm 1}United Nations University Institute in Macau, Macau SAR, China\\
    \textsuperscript{\rm 2}AI Singapore\\
    \textsuperscript{\rm 3}Data Science for Social Impact Lab, University of Pretoria, South Africa\\
    \textsuperscript{\rm 4}Data Science for Social Impact Lab \& African Institute for Data Science and AI, University of Pretoria, South Africa
}

\begin{document}
\maketitle

\begin{abstract}
Public policies addressing urgent social and environmental challenges need to explicitly consider the diverse, often conflicting perspectives of the affected stakeholders. Despite computational decision-support approaches increasingly offering recommendations across diverse human value systems, they still tend to deliver a single consensus-driven outcome. We present \sysname{}, a modular system in which multiple large language model agents, each grounded in distinct demographic stakeholder personas and conceptual frameworks, deliberate on policy scenarios.
The objective of \sysname{} is not merely to inform about a preferred policy proposal, but also to automatically surface who is affected by the scenario and provide decision-makers with the rationales and quantitative indicators behind each position. \sysname{} treats non-convergence as a first-class explainable output, enabling policymakers to better understand the trade-offs and human impacts of the policies being discussed.
\end{abstract}

\section{Introduction}
General-purpose structured multi-agent debate has primarily served as a method for alignment, but its potential to explore disagreement to support human decision-making remains largely unutilized. The original multi-agent debate framework \citep{du_improving_2024} and later work disentangling debate from voting \citep{choi_debate_2026,kaesberg_voting_2025} treat divergence among agents as a topic to be resolved by majority vote. In turn, the Habermas Machine \citep{tessler_ai_2024} aims to converge deliberately by iteratively aligning with majoritarian human points of view.
While offering a single recommendation that aligns with main stream human value systems is a desirable outcome from general purpose multi-agent LLM systems, the complexity and high stakes involved in public policymaking requires careful examination of the rationale and quantitative impact of the often confronted stakeholders' points of view.

Within the specific domain of multi-agent LLM systems for pluralistic deliberation, ADEPT \citep{zohny_simulating_2025} presents a debate among ethicists over a fixed medical-triage scenario and reports the diversity of perspectives. Similarly, \citet{steging_investigating_2026} presents persona-grounded architectures for legal topics answering via courtroom-inspired role assignment. Demographically conditioned LLM personas have also been used to predict or replicate observed legislative behavior, e.g., roll-call voting \citep{li_political_2024} and European Parliament voting patterns \citep{kreutner_persona-driven_2026}.

However, to the best of our knowledge, no existing multi-agent LLM system combines: (i) grounding agents in both demographic identity and explicit ethical conceptual frameworks, (ii) treating non-convergence as a reportable outcome in its own right, and (iii) allowing users to choose between multiple indicators against which to evaluate the deliberation's outcome. \sysname{} does all three, through a pipeline of installable plugins: it builds a stakeholder roster, runs a structured debate in which disagreements are stated explicitly, and reports the potential outcomes (for each position) under several value systems. \sysname{}\footnote{The name \emph{KITA AI} corresponds to the inclusive form of ``we/us'' in Indonesian, and it was chosen in recognition of the needs voiced by the country's policymakers.} has been designed according to the needs expressed by UN stakeholders to the United Nations University.

\section{System Overview}
\begin{figure}[t]\centering
  \resizebox{0.84\linewidth}{!}{
%
%
\definecolor{pcInk}{HTML}{1F2933}%
\definecolor{pcRole}{HTML}{2C5282}%
\definecolor{pcOpt}{HTML}{6B7280}%
\definecolor{pcJson}{HTML}{8A5A00}%
\definecolor{pcCtx}{HTML}{0E7C7B}%
\def\icn#1{\textcolor{pcCtx!80!black}{#1}}%
\def\ico#1{\textcolor{pcOpt!50}{#1}}%
\def\icoo#1{\textcolor{pcOpt!35}{#1}}%
\def\icsep{\hspace{4pt}}%
\def\xl{0}\def\xr{4.65}%
\def\ya{0}\def\yb{-2.05}\def\yc{-4.1}\def\yd{-6.15}%
\begin{tikzpicture}[
  font=\sffamily,
  >={Stealth[length=2.6mm,width=2.4mm]},
  stage/.style={rectangle, rounded corners=3pt, draw=pcRole, line width=0.8pt,
    fill=pcRole!7, text width=4.0cm, minimum height=0.8cm, align=center,
    inner xsep=3pt, inner ysep=2pt, text=pcInk, font=\sffamily\small\bfseries},
  stageopt/.style={stage, draw=pcOpt, fill=pcOpt!6, dash pattern=on 4pt off 2.5pt},
  artifact/.style={stage, rounded corners=1.5pt, draw=pcJson, fill=pcJson!10,
    font=\ttfamily\small},
  outfile/.style={stage, draw=pcOpt, line width=0.7pt, fill=pcOpt!8,
    font=\ttfamily\small},
  chip/.style={rectangle, rounded corners=2.5pt, draw=pcCtx!55, line width=0.7pt,
    fill=pcCtx!8, inner xsep=6pt, inner ysep=3.5pt, font=\small},
  flow/.style={draw=pcCtx!65!black, line width=1pt, rounded corners=4pt},
  flowin/.style={flow, ->},
  ctxnote/.style={font=\sffamily\small\itshape, text=pcCtx!70!black,
    inner sep=1pt, anchor=east},
]
\node[artifact] (srj) at (\xl,\ya) {%
  {\normalfont\large\color{pcJson!85}\faQuestion}\,\textbf{\{\,\}}\ request.json};
\node[stageopt] (ctx) at (\xl,\yb) {%
  {\normalfont\large\color{pcOpt!85}\faGlobe}\ \ Contextualiser};
\node[stage] (ab) at (\xl,\yc) {%
  {\normalfont\large\color{pcRole!85}\faUsers}\ \ Agent Builder};
\node[stage] (de) at (\xl,\yd) {%
  {\normalfont\large\color{pcRole!85}\faComments}\ \ Debate Engine};
\node[stageopt] (sim) at (\xr,\yc) {%
  {\normalfont\large\color{pcOpt!85}\faChartLine}\ \ Simulation};
\node[stage] (al) at (\xr,\yb) {%
  {\normalfont\large\color{pcRole!85}\faBalanceScale}\
  \mbox{Report Engine}\,$\times$N};
\node[outfile] (rep) at (\xr,\ya) {%
  {\normalfont\large\color{pcInk!75}\faFileCode}\ \ report.html};
\node[chip] (c1) at ($(srj.south)!0.5!(ctx.north)$) {\icn{\faQuestion}};
\node[chip] (c2) at ($(ctx.south)!0.5!(ab.north)$)
  {\ico{\faQuestion}\icsep\icn{\faGlobe}};
\node[chip] (c3) at ($(ab.south)!0.5!(de.north)$)
  {\ico{\faQuestion}\icsep\icoo{\faGlobe}\icsep\icn{\faUsers}};
\node[chip] (c4) at (\xr,\yd)
  {\ico{\faQuestion}\icsep\icoo{\faGlobe}\icsep\ico{\faUsers}\icsep\icn{\faComments}};
\node[chip] (c5) at ($(sim.north)!0.5!(al.south)$)
  {\ico{\faQuestion}\icsep\icoo{\faGlobe}\icsep\ico{\faUsers}\icsep
   \ico{\faComments}\icsep\icn{\faChartLine}};
\node[chip] (c6) at ($(al.north)!0.5!(rep.south)$)
  {\ico{\faQuestion}\icsep\icoo{\faGlobe}\icsep\ico{\faUsers}\icsep
   \ico{\faComments}\icsep\icoo{\faChartLine}\icsep\icn{\faBalanceScale}};
\begin{scope}[on background layer]
  \draw[stage, fill=pcRole!4]
    ($(al.north west)+(0.12,0.12)$) rectangle ($(al.south east)+(0.12,0.12)$);
  \draw[stage, fill=pcRole!5]
    ($(al.north west)+(0.06,0.06)$) rectangle ($(al.south east)+(0.06,0.06)$);
\end{scope}
\draw[flow] (srj.south) -- (c1.north);  \draw[flowin] (c1.south) -- (ctx.north);
\draw[flow] (ctx.south) -- (c2.north);  \draw[flowin] (c2.south) -- (ab.north);
\draw[flow] (ab.south)  -- (c3.north);  \draw[flowin] (c3.south) -- (de.north);
\draw[flow] (de.east)   -- (c4.west);   \draw[flowin] (c4.north) -- (sim.south);
\draw[flow] (sim.north) -- (c5.south);  \draw[flowin] (c5.north) -- (al.south);
\draw[flow] (al.north)  -- (c6.south);  \draw[flowin] (c6.north) -- (rep.south);
\node[ctxnote] at ($(c1.west)+(-0.12,0)$) {transcript};
\end{tikzpicture}}
  \caption{Architecture of \sysname{}.}
  \label{fig:arch}
\end{figure}

The architecture of \sysname{} consists of three mandatory modules that run sequentially, plus two optional ones.
Every module is defined by what it receives and what it returns, not by how it works internally. Specific implementations of the module task become \textit{plugins} of that module.
As shown in Figure~\ref{fig:arch}, the system incrementally builds a transcript artifact that traverses the entire pipeline, providing the modules with the context they need for their tasks.

\textbf{The Agent Builder} (AB) maps a policy scenario to a roster of agent personas or stakeholders. Representing a diversity of conceptual positions is not trivial. Since a missing voice cannot be recovered downstream, the selected roster must include at least the prominent conflicting voices in the policy challenge. Our implemented plugin lets an LLM propose the relevant demographic axes and the value system anchoring each persona, or takes both directly from the user.

\textbf{The Debate Engine} (DE) takes the policy scenario and the agents from the AB and generates an argument graph induced by the debate's moves. DE aims to make disagreement explicit, resolve it if possible, or otherwise characterize it. Our plugin casts each proposal in Toulmin form \citep{toulmin_uses_1958} and each round moves as a formal dialogue game \citep{prakken_formal_2006}. An objection quotes its target argument and is conceded if left unanswered. Consensus is then derived from the argument graph, in which a proposal survives only if the contradictory arguments are aligned \citep{dung_acceptability_1995}.

\textbf{The Report Engine} maps the transcript of the policy proposals resulting from the debate to a report, reading each candidate's proposals through their own value-system lens. Any number of engines can be run concurrently on the same transcript, each producing its own report. We publish a set of value-systems engines that comprise a diversity of established philosophical traditions and widely accepted indicators, including the UN Sustainable Development Goals.

\textbf{Optional Modules.} The optional modules supply context beyond the policy scenario description alone.
The \emph{Contextualizer} runs before the AB, browsing the internet for information from relevant sources to supplement the initial scenario description. Other uses of this module could include gathering data from identified databases, performing statistical analysis on provided data, or consulting selected debate forums for a fuller picture of the public opinion.

The \emph{Simulator} module aims to leverage external social models, such as agent-based models \citep{bonabeau2002agent} or society-of-agents models \citep{li_camel_2023}, to evaluate the impact of policy implementation. It is therefore highly domain-specific, as policy scenarios might require tailored simulations in order to be reliable. Some examples of plausible plugins include the works of \citet{park_generative_2023} for social simulation, \citet{sultimov_respond_2026} for disaster response, and \citet{li_agent_2024} for hospital simulation.

\section{Modularity and Scalability}

\sysname{}'s architecture is designed to be replicable and scalable. First, it is LLM-agnostic by design, routing every agent call through OpenRouter, so as new models become faster, cheaper, or more capable, \sysname{} can adopt them without a redesign. Second, every module reads and writes to a single shared \textit{transcript} rather than calling adjacent modules directly, so modules can be developed, tested, and swapped independently of whatever sits upstream or downstream. Third, because modules are built as installable plugins constituting a library, \sysname{} is designed to grow as a community project, with enhanced search wrappers, region-specific report engines, or different debate engines becoming installable behaviors. To facilitate contributions, each module includes a template with an \texttt{agents.md} file guiding the creation of new plugins.

\section{Use Cases and User Interface}

The modular design allowed us to test three different use cases. The default pipeline takes a policy scenario and produces and analyzes policy proposals. The backtesting pipeline uses a Contextualizer plugin that restricts web searches to dates before the conclusion of a policy debate to reproduce the outcome. Finally, the perturbation pipeline extends backtesting to mitigate data leakage from training data. It reproduces past scenarios by changing its initial formulation and checks whether the system adapts the proposals accordingly.
A full pass through the three pipelines takes several minutes, so a set of precomputed runs is available through a user-friendly UI. The tool also includes a \textit{Module Lab} for testing specific modules with manual input. Through the interface, users can propose a brand-new policy for a given scenario, view the evaluation, observe the outcomes of discussions, run debates between predefined personas on specific topics, or build a plugin by talking to a coding agent.

\section{Responsible Use and Limitations}
\sysname{} is a deliberation-support tool, not a decision procedure. The generated deliberation carries the biases of the underlying models even when
personas are explicitly conditioned through prompting. Consequently, \sysname{} outcomes should not be regarded as consultations with the population, and the rationales must be interpreted by domain experts rather than trusted directly.

\section{Conclusion}
\sysname{} is a novel tool that leverages diverse stakeholders' perspectives on a social challenge and offers multi-faceted viewpoints and indicators that support decision makers. Its open, modular design allows for plugins, scalability, and customization to region-specific value systems.

\section{Acknowledgments}
This research is supported by the National Research Foundation, Singapore under its National Large Language Models Funding Initiative. Any opinions, findings and conclusions or recommendations expressed in this material are those of the author(s) and do not reflect the views of National Research Foundation, Singapore. The authors acknowledge the Data Science for Social Impact (DSFSI) Lab, supported by the ABSA Chair of Data Science and the African Institute of Data Science and AI (AfriDSAI). This work was supported by the AI4D Africa Program (African Languages Lab), funded by the International Development Research Centre (IDRC), Canada and Foreign, Commonwealth \& Development Office (FCDO), UK.

\bibliography{references}
\end{document}